\documentclass[a4paper,aip,reprint]{revtex4-1}
\usepackage{graphicx}
\usepackage{amsmath}
\usepackage{fancyhdr}
\usepackage{amssymb}
\usepackage{color}
\usepackage{array}
\usepackage{booktabs}
\begin{document}
\title{TMI: Thermodynamic inference of data manifolds}
\author{Purushottam D. Dixit}
\thanks{email:pdixit@ufl.edu}
\affiliation{Department of Physics, University of Florida, Gainesville, FL, United States}
\begin{abstract}
The Gibbs-Boltzmann distribution offers a physically interpretable way to massively reduce the dimensionality of high dimensional probability distributions where the extensive variables are `features' and the intensive variables are `descriptors'. However, not all probability distributions can be modeled using the Gibbs-Boltzmann form.  Here, we present TMI: TMI, {\bf T}hermodynamic {\bf M}anifold {\bf I}nference; a thermodynamic approach to approximate a collection of arbitrary distributions. TMI simultaneously learns from data intensive and extensive variables and  achieves dimensionality reduction through a multiplicative, positive valued, and interpretable decomposition of the data. Importantly, the reduced dimensional space of intensive parameters is not homogeneous. The Gibbs-Boltzmann distribution defines an analytically tractable Riemannian metric on the space of intensive variables allowing us to calculate geodesics and volume elements.  We discuss the applications of TMI with multiple real and artificial data sets. Possible extensions are discussed as well. 
\end{abstract}
\maketitle


{\bf Introduction:} Scientific data often comprise positive numbers. Examples include pixels intensities of grayscale images~\citep{lecun2010mnist}, abundances of bacteria in microbial ecosystems~\citep{ji2019quantifying}, electrical activities of brain regions~\citep{saxena2019localized}, or more generally a collection of probability distributions; all of whom once suitably normalized can be manipulated as probability distributions.

Over the past few years, our ability to collect high quality high dimensional data has improved substantially which has been accompanied by a flurry of dimensionality reduction methods. These  methods usually belong to one of two  broad classes. Methods such as principal component analysis (PCA), singular value decomposition (SVD), and non-negative matrix factorization (NMF)~\citep{lee1999learning,hofmann1999probabilistic} are examples of matrix factorization based methods. Here, the high dimensional data (in the form of a matrix) is expressed as a multiplication of two or more {\it simpler} (for example, sparse or low rank) matrices. In contrast methods such as diffusion maps~\citep{coifman2006diffusion},  Laplacian Eigenmaps~\citep{belkin2003laplacian}, Isomaps~\citep{balasubramanian2002isomap}, tSNE (t-stochastic neighborhood embedding)~\citep{maaten2008visualizing}, and UMAP (uniform manifold approximation and projection)~\citep{mcinnes2018umap} are based on manifold learning. These methods rely on the assumption that the high dimensional data lies on a much lower dimensional embedded manifold. These methods infer the manifold using estimation of local density of data points in the higher dimensions using kernel based approaches.  

Orthogonal to these modern approaches, statistical physics offers a physically interpretable solution to dimensionality reduction; albeit for a restricted class of distributions. Consider a system at thermodynamic equilibrium with a surrounding bath that can exchange $K-$types of extensive variables with it. Let the number of states in the system be $d$.  Typically, $d \gg 1$ ($d\sim 10^{23}$ for a mole of ideal gas) and $K\sim o(1) \ll d$ ($K = 1, 2$ for the canonical and the grand canonical ensemble respectively). Imagine that there are $N$ different realizations of the bath; characterized by $N\times K$ Lagrange mulitpliers $\lambda_k^{(\alpha)}$ ($k\in [1, K]$, $\alpha \in [1, N]$). At thermodynamic equilibrium, any realization  $\alpha$ can be described by the $K$ intensive variables; the probability $q_{a}^{(\alpha)}$ of observing the system in state `$a$' is given by the Gibbs-Boltzmann distribution:
\begin{eqnarray}
q_{a}^{(\alpha)} = \frac{1}{Z^{(\alpha)}} \exp \left ( -\sum_{k=1}^{K} \lambda_{k}^{(\alpha)}Y_{ka} \right) =  \frac{1}{Z^{(\alpha)}}\prod_k  \left (\gamma_{k}^{(\alpha)} \right )^{Y_{ka}}\label{eq:gb}
\end{eqnarray}
In Eq.~\ref{eq:gb}, $\lambda^{(\alpha)}_{k}$ are realization-specific Lagrange multipliers, $Y_{ka}$ are state-dependent extensive variables, 
\begin{eqnarray}
Z^{(\alpha)} &=&\sum_a\exp \left ( -\sum_{k=1}^{K} \lambda_{k}^{(\alpha)}Y_{ka}\right ) \label{eq:pf}
\end{eqnarray}
is the partition function, and
\begin{eqnarray}
\gamma_{k}^{(\alpha)} = \exp \left (-\lambda_{k}^{(\alpha)} \right) > 0
\end{eqnarray}
are generalized activity coefficients. Importantly, recent work has shown that the Gibbs-Boltzmann form has a much broader applicability, even beyond thermal systems at thermodynamic equilibrium. Notably, using the maximum entropy principle~\citep{dixit2018perspective}, it has been employed to model probabilities in a variety of complex systems such as ensembles of protein sequences~\citep{cocco2018inverse}, collective firing of neurons~\citep{savin2017maximum}, and collective motions of birds~\citep{bialek2012statistical}.

Unfortunately, however, not every collection $\{ {\bf x}^{(\alpha)} \}, \alpha\in [1, N]$ of $N$ abitrary probability distributions can be described using the exponential Gibbs-Boltzmann form. Here, we ask the following question: Given data in the form of $N$ arbitrary distributions $\{ {\bf x}^{(\alpha)} \}$, can we infer approximate extensive variables $Y$s and intensive variables $\lambda$s such that the Gibbs-Boltzmann form in Eq.~\ref{eq:gb} approximates the given distributions $\{ {\bf x}^{(\alpha)} \}$?

To that end, we introduce TMI: {\bf T}hermodynamic {\bf M}anifold {\bf I}nference. In TMI, we simultaneously infer from the data extensive variables (`energies') and intensive variables (`temperatures'). The extensive variables represent features on the state space while the intensive variables embed the data points in a lower dimensional space. TMI achieves several objectives. First, by enforcing the number of extensive variables to be much smaller than the data dimension, it achieves dimensionality reduction. Notably, unlike principal component analysis (PCA) or singular value decomposition, but similar to non-negative matrix factorization~\citep{lee1999learning,hofmann1999probabilistic}, TMI-based approximation of the data leads to interpretable positive-valued factorization (see Eq.~\ref{eq:gb}). Second, TMI defines a Riemannian manifold with an analytically tractable distance metric on the space of intensive variables where the data points reside.  Importantly, this metric allows us to define geodesic distances between arbitrary points in the space of intensive variables as well as volume elements. Third, due to the convexity of the Gibbs-Boltzmann equation, TMI provides a unique out-of-sample extension~\citep{bengio2004out} procedure. We illustrate TMI using several real and artificial datasets. 

{\bf TMI approximates arbitrary distributions:} Consider data in the form of discrete distributions  $\{ {\bf x}^{(\alpha)}\}, \alpha \in [1, N]$ defined on a $d$ dimensional state space. We assume that ${\bf x}_{a}^{(\alpha)} > 0~\forall~a \in [1,d]$ and $\forall~\alpha \in [1, N]$. We want to find $K$ $d-$dimensional extensive variables $\{ \bar Y_k \} \equiv \{ Y_{ka} \}$ and $N$ $K-$dimensional intensive bath parameters $\{ \bar \lambda^{(\alpha)} \} \equiv \{ \lambda_{k}^{(\alpha)} \}$ such that the Gibbs-Boltzmann distributions in Eq.~\ref{eq:gb} approximate the original distributions ${\bf x}^{(\alpha)}$.

In TMI, we enforce $K \ll N$ to obtain an approximate lower dimensional representation of each distribution. To that end,  for a fixed $K$, we minimize the sum of Kullback-Leibler divergences between ${\bf x}^{(\alpha)}$ and $q^{(\alpha)}$:
\begin{eqnarray}
C &=& \sum_\alpha \sum_a {\bf x}_{a}^{(\alpha)} \log \frac{{\bf x}_{a}^{(\alpha)}}{q_{a}^{(\alpha)}}
\end{eqnarray}
The first term in the expanded KL divergence depends only on the distributions ${\bf x}^{(\alpha)}$ and can be dropped. We have
\begin{eqnarray}
C &=& -\sum_\alpha \left ( \sum_a {\bf x}_{a}^{(\alpha)} \log q_{a}^{(\alpha)} \right ) \\
&=&\sum_\alpha \left ( \sum_a {\bf x}_{a}^{(\alpha)}\left ( \sum_{k=1}^{K} \lambda_{k}^{(\alpha)}Y_{ka} + \log Z^{(\alpha)} \right ) \right )\\
&=&\sum_\alpha \log Z^{(\alpha)} + \sum_{\alpha, a, k}  {\bf x}_{a}^{(\alpha)}\lambda_{k}^{(\alpha)} Y_{ka}\label{eq:c0}
\end{eqnarray}

There are several indeterminacies in the cost function in Eq.~\ref{eq:c0}. First, for a fixed $k$, the cost is invariant to to an additive shift $Y_{ka} = Y_{ka} + c~\forall~a\in [1, d]$.  This corresponds to the translational invariance in energies in a physical system. Second, the cost is invariant with respect to a scaling $\lambda_{k}^{(\alpha)} \rightarrow B\times \lambda_{k}^{(\alpha)} $ for all distributions $\alpha \in [1, N]$ and a corresponding transformation that scales $Y_{ka} = Y_{ka}/B$ for all $a\in [1,d]$. Physically, this corresponds to the fact that extensive variables (for example, energies) are always multiplied by the corresponding intensive variables (for example, inverse temperatures) when computing probabilities. More generally, if we multiple the $d\times K$ matrix of extensive variables by a $K\times K$ matrix {\bf B} and multiple the $N\times K$ matrix of intensive variables with $\left ({\bf B}^{-1} \right)^{\rm T}$, the Gibbs-Boltzmann probabilities  don't change. Finally, the cost  is invariant to permutations in  $k$, the label of the extensive variables.  

We resolve the first indeterminacy by first finding a converged set of variables $Y$s and then setting the lowest one to zero.  We resolve the second indeterminacy by constraining the $L_2$ norm of the extensive variables. We do this by introducing constraints in the cost  using Lagrange multipliers. The modified cost function is given by
\begin{eqnarray}
C = \sum_\alpha \log Z^{(\alpha)} + \sum_{\alpha, a, k}  {\bf x}_{a}^{(\alpha)}\lambda_{k}^{(\alpha)} Y_{ka} + \sum_k \beta_k \left ( \sum_a Y_{ka}^2 \right ) \nonumber \label{eq:cost_mod} \\
\end{eqnarray}
Finally, we resolve the third indeterminacy by rank-ordering the parameters $\lambda_{k}^{(\alpha)}$ by their $L_2$ norm across all samples. 

The cost  is convex respect to $\lambda$s when $Y$s are fixed and vice versa. However, similar to non-negative matrix factorization~\citep{lee1999learning}, it is not guaranteed to be globally convex (see appendix~\ref{ap_convex}). We can minimize $C$ with respect to the intensive and the extensive variables to find a local minimum. Differentiating with respect to $\lambda_{k}^{(\alpha)}$ and setting the derivative to zero,
\begin{eqnarray}
\sum_a q_{a}^{(\alpha)} Y_{ka} = \sum_a {\bf x}_{a}^{(\alpha)}Y_{ka}.~\forall~k\in [1, K].\label{eq:dl}
\end{eqnarray}
Eq.~\ref{eq:dl} has a simple interpretation: when the values of the extensive variables $Y$ are fixed, the intensive variables $\lambda_{k}^{(\alpha)}$ describing any particular distribution are determined by matching the averages of the extensive variables predicted using $q^{(\alpha)}$ with their empirical average values computed using the actual distributions ${\bf x}^{(\alpha)}$.

For a fixed value of $\lambda$s, the value of $Y_{ka}$ are the fixed points of a non-linear equation. Differentiating $C$ with respect to $Y_{ka}$ and setting the derivative to zero,
\begin{eqnarray}
0 &=& -\sum_{\alpha} \lambda_{k}^{(\alpha)}q_{a}^{(\alpha)} + 2\beta_k Y_{ka} + \sum_{\alpha}  {\bf x}_{a}^{(\alpha)}\lambda_{k}^{(\alpha)} \\
\Rightarrow Y_{ka} &=& \frac{1}{2\beta_k}\left ( \sum_{\alpha}\lambda_{k}^{(\alpha)} \left ( {q}_{a}^{(\alpha)}   -{\bf x}_{a}^{(\alpha)} \right )  \right ) \label{eq:dy}
\end{eqnarray}
Note that both sides of Eq.~\ref{eq:dy} depend on $Y_{ka}$ since the distributions $q^{(\alpha)} $ depend on $Y_{ka}$. 

Above, for any fixed $k$, the inference of the extensive variables $Y_{ka}$s is invariant to a permutation over labeling of the state space indices $\{ a \}$. However, it is possible to incorporate information about the geometrical structure of the state space in the inference as well. One such structure is smoothness. Consider the example of grayscale images. Here, the distributions represent normalized pixel intensities of a digitized image. In the images, any state `$a$' is identified by planar two dimensional coordinates $a\equiv (i, j)$ which define adjacency in the state space. Let us consider two adjacent states $a\equiv (i, j)$ and $b\equiv a + \hat e$ ($\hat e \in \{ (1,0), (0,1), (-1, 0), (0, - 1) \}$). We can ensure that the extensive variables $Y_{ka}$ and $Y_{kb}$ corresponding to neighboring states `$a$' and `$b$' are similar to each other by introducing regularizing constraints:
\begin{eqnarray}
\sum_{a,b} n_{ab} \left (Y_{ka} - Y_{kb}\right)^2 < C_k~\forall~k \label{eq:smooth}
\end{eqnarray}
where $n_{ab} = 1$ when $a$ and $b$ are adjacent and zero otherwise. Such constraints will limit the {\it ruggedness} of the landscape of the extensive variables. Other constraints on the extensive variables, such as orthogonality, can also be imposed.

Similarly, constraints on the intensive variables can be imposed as well. The formulation developed above will lead to intensive variables that are both positive and negative. However, while the TMI-based factorization of the data still remains positive, nonnegativity constraints on the intensive parameters $\lambda$s may be desirable in order to interpret the extensive variables as potential energy minima. These can be indirectly imposed by employing the multiplicative update algorithm, as is done in nonnegative matrix factorization~\citep{lee1999learning}, to infer $\lambda$s as opposed to a gradient or Hessian descent algorithm. The details of the numerical algorithms to learn both $Y$s and $\lambda$s are in appendix~\ref{ap_numerical}.  

Finally, we note that though the above discussion was restricted to data in the form of normalized distributions, TMI can also be implemented to unnormalized positive valued data. Notably, the equations to determine $Y$s and $\lambda$s are identical to those presented above (Eq.~\ref{eq:dl} and Eq.~\ref{eq:dy}). We present this development in detail in appendix~\ref{ap_unnormalized}.

{\bf TMI provides a unique out-of-sample extension procedure:} A common situation in data analysis is as follows. Suppose that we have inferred $\lambda$s and $Y$s from $N$ data points using TMI. Now imagine that a $N+1$st data point arrives. Can we approximately embed this data point in the lower dimensional space? This problem is commonly known as the out-of-sample extension~\citep{bengio2004out} and usually does not have a unique solution~\citep{coifman2006geometric}. 

Notably, in TMI a new data point ${\bf x}^{(\nu)} $ can be embedded rapidly by determining the $K$ Lagrange multipliers $\lambda_{k}^{(\nu)} $ by solving for $\bar \lambda^{(\nu)}$:
\begin{eqnarray}
\sum_a q_{a}^{(\nu)} (\bar \lambda^{(\nu)}) Y_{ka} = \sum_a x_{a}^{(\nu)} Y_{ka}~\forall~k \in [1, K].
\end{eqnarray}
Moreover, the quality of the embedding can be assessed by evaluating the KL divergence
\begin{eqnarray}
KL = \sum_a x_{a}^{(\nu)}  \log \frac{x_{a}^{(\nu)} }{q_{a}^{(\nu)} }.
\end{eqnarray}

{\bf TMI introduces a Riemannian distance metric:} Several functionals can quantify the differences between distributions. These include traditional quantifiers such as the Kullback-Leibler divergence~\citep{kullback1951information}, Bhattacharya distance~\citep{bhattacharyya1943measure}, and Hellinger distance~\citep{nikulin2001hellinger} which  are invariant with respect to permutations of state space indices. In contrast, the optimal transport distance  (also known as the Wasserstein distance)~\citep{peyre2017computational,amari2018information,amari2019information} is a distance metric that takes into account the geometry of the state space.

TMI defines a Riemannian geometry and a distance metric on the space of intensive variables. Consider two different distributions approximated by intensive parameters $\bar \lambda^{(1)} $ and $\bar \lambda^{(2)} $. Consider a smooth and differentiable path $\gamma(t)$ between the two distributions such that $\gamma(t = 0) = \bar \lambda^{(1)} $ and $\gamma(t=T) = \bar \lambda^{(2)} $. In the linear response regime, the excess work -- work done above the difference in thermodynamic potentials --  along this path can be computed~\citep{crooks2007measuring,sivak2012thermodynamic,rotskoff2015optimal,rotskoff2017geometric}:
\begin{eqnarray}
P \propto \int \limits_{0}^{T}  \frac{d\bar \lambda^{\rm T}}{dt}g(\bar \lambda) \frac{d\bar \lambda}{dt}dt
\end{eqnarray}
where the elements of the friction tensor $g$ are given by~\citep{sivak2012thermodynamic}
\begin{eqnarray}
g_{ij}(\bar \lambda) = \int \limits_{0}^{\infty} \langle\delta Y_i (0) \delta Y_j(\tau) \rangle_{\bar \lambda} d\tau \label{eq:gij}
\end{eqnarray}
In Eq.~\ref{eq:gij}, $\delta Y_i = Y_i - \langle Y_i\rangle$ where $\langle Y_i\rangle$ is the ensemble average value of the extensive variable $Y_i$ when the intensive parameters are fixed at $\bar \lambda(t)$. We note that a similar derivation exist for transforming two non-equilibrium steady state (NESS) distributions~\citep{mandal2016analysis}. However, NESS distributions cannot be expressed in the parametric Gibbs-Boltzmann form and therefore we do not pursue that direction here. 

The friction tensor depends on the dynamics on the state space $\{ a \}$ at a fixed $\bar \lambda$. When the transition rate matrix $\kappa_{a\rightarrow b}(\bar \lambda)$ is provided, the friction tensor can be computed in a straightforward manner (see appendix~\ref{ap_friction}). What are reasonable choices for the dynamics? We want an `equilibrium' (detailed balanced) transition rate matrix that is constrained to reproduce the Gibbs-Boltzmann distribution $q(\bar \lambda)$. One way to incorporate the information about the underlying geometry is to require that the rates  penalizes transitions between geometrically `distant' states $a$ and $b$. A simple transition rate matrix is the one that maximizes the path entropy~\citep{dixit2015inferring}: 
\begin{eqnarray}
\kappa_{a\rightarrow b}(\bar \lambda) \propto \sqrt{\frac{q_{b}(\bar \lambda)}{q_{a}(\bar \lambda)}}\exp \left ( -\frac{d(a,b)^2}{\varepsilon}\right ). \label{eq:kappa}
\end{eqnarray} 
Another choice for the dynamics is the so-called Glauber dynamics~\citep{glauber1963time}:
\begin{eqnarray}
\kappa_{a\rightarrow b}(\bar \lambda) \propto \frac{q_{b}(\bar \lambda)}{q_{a}(\bar \lambda)+q_{b}(\bar \lambda)}\exp \left ( -\frac{d(a,b)^2}{\varepsilon}\right ). \label{eq:kappa_g}
\end{eqnarray} 
In Eq.~\ref{eq:kappa} and Eq.~\ref{eq:kappa_g}, $d(a,b)$ is a measure of separation between states $a$ and $b$ (for example Euclidean distance) and $\varepsilon > 0$ plays the role an inverse diffusion constant. Finally, we note any choice of the dynamics will define a well-behaved friction tensor that as long as the dynamics is reversible and reproduces the stationary distribution $q(\bar \lambda)$. 

From the dynamics, the friction tensor can be calculated in a straightforward manner as shown in Appendix~\ref{ap_friction}. The distance  computed using this friction tensor will be a proper distance metric which respects the underlying geometry of the state space. We note that unlike the Wasserstein distnce, TMI defines a distance metric even when the measure $d(a,b)$ is not a proper distance metric. Moreover, a significant advantage of this geodesic approach  is that it can be used to compute an optimal path of transition for a pair of intensive variables.

Notably, when the dynamics is fast, the friction coefficient reduces (up to a proportionality) to the Fisher information matrix~\citep{crooks2007measuring,sivak2012thermodynamic,rotskoff2015optimal,rotskoff2017geometric}, which in the case of Gibbs-Boltzmann distributions is the matrix of fluctuations~\citep{caticha2008lectures}.  Moreover, if we assume that the rate of change of $\bar \lambda$ along a trajectory is kept constant, the paths that minimize excess work are also the paths that minimize the geodesic distance~\citep{crooks2007measuring,sivak2012thermodynamic,rotskoff2015optimal,rotskoff2017geometric}. Hence, the length of the path of minimum excess work between two distributions, described by $\bar \lambda_1$ and $\bar \lambda_2$ respectively, also defines a metric distance  between them. We note however that the Fisher information matrix is invariant to a permutation of the indices. Therefore,  the geodesic distances evaluated using the Fisher information matrix does in itselt not take into account the geometry of the state space. 

Finally, we note that the distance metric is defined on the space of intensive variables and not the distributions themselves.

\begin{figure}
       \begin{center}
               \includegraphics[scale=0.38]{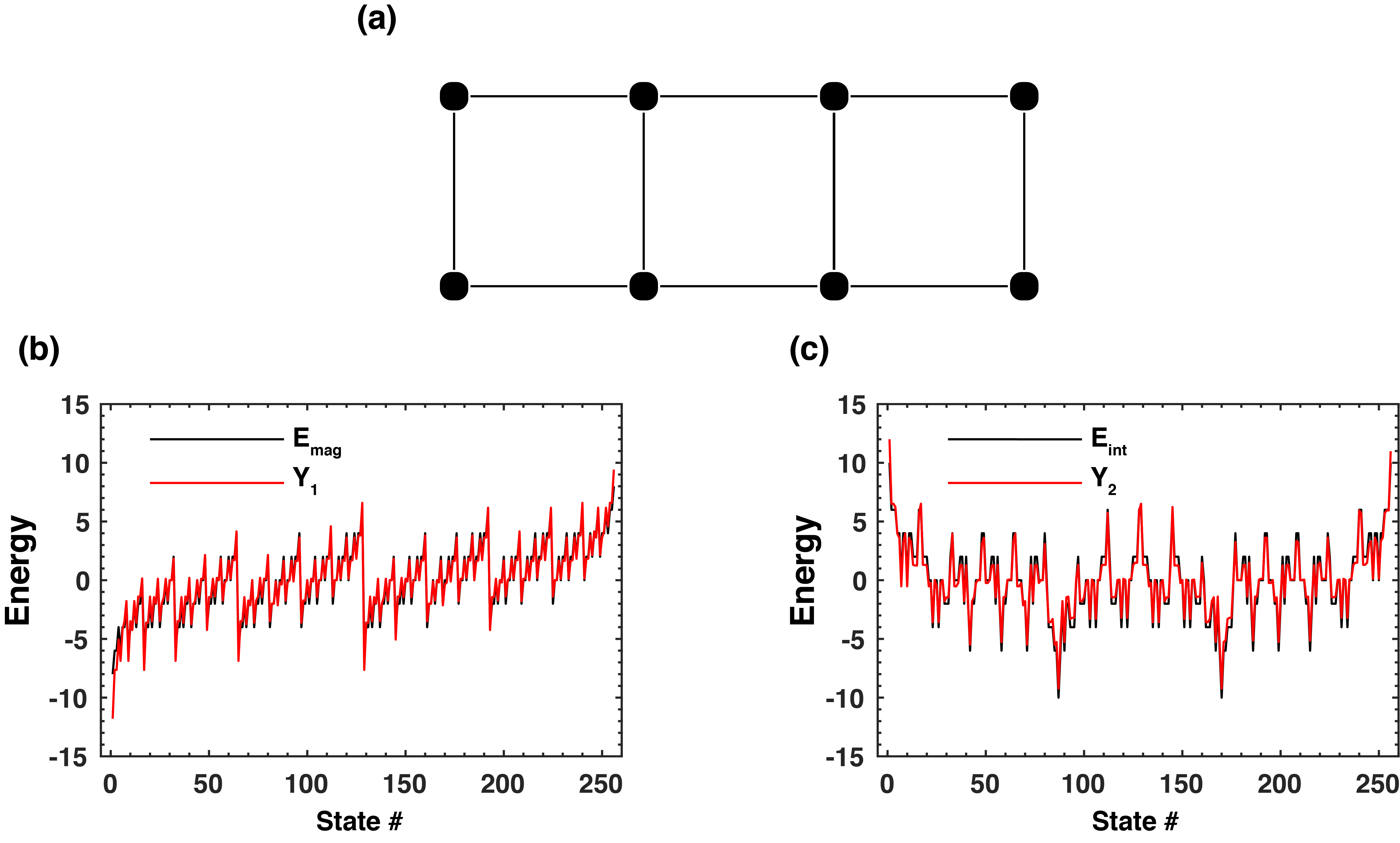}
               \caption{{\bf panel (a)} the connectivity graph of a 8 spin Ising model, {\bf panel (b)} inferred extensive variable $\bar Y_1$ (red) compared to the true extensive variable $E_{\rm mag}$ (black), and {\bf panel (c)} inferred extensive variable $\bar Y_2$ (red) compared to the true extensive variable $E_{\rm int}$ \label{fg:ising_comparison}}
       \end{center}
\end{figure}

{\bf Learning Ising model from data}: As a test case, we show that TMI can infer the energy landscape of an Ising model from sampled distributions. We consider a nearest-neighbor Ising model with $n_s = 8$ spins arranged as shown in panel (a) of Fig.~\ref{fg:ising_comparison}. Each spin $\sigma$ can take values 1 or $-1$. The probability of observing any spin configuration $\bar \sigma(a)$ is given by
\begin{eqnarray}
p(\bar \sigma(a)) = \frac{1}{Z(H,J)}\exp \left ( -H  E_{\rm mag}(a)  - J E_{\rm int}(a) \right )\label{eq:ising_energy}
\end{eqnarray}
where
\begin{eqnarray}
E_{\rm mag}(a) &=& \sum_{i} \sigma(a)_i,~{\rm and}\\
E_{\rm int}(a) &=& \sum_{i~{\rm nn}~j}\sigma(a)_i \sigma(a)_j.\label{eq:int}
\end{eqnarray}
In Eq.~\ref{eq:int}, the summation is taken over the nearest neighbors of the graph shown in panel (a) of Fig.~\ref{fg:ising_comparison} and $Z(H,J)$ is the partition function.

We  randomly sampled 50 pairs of $H$ and $J$ values from a uniform distribution where $H \in [-1, 1]$ and $J\in [-1, 1]$ and generated 50  Ising model distributions (see Fig.~\ref{fg:ap_ising}). Next, we approximated these input distributions using TMI with $K=2$ extensive variables  $\bar Y_1$ and $\bar Y_2$. We simultaneously inferred  50 pairs of Lagrange multipliers  representing each of the 50 distributions.

As noted above, multiplication by a matrix $Y\rightarrow Y \times {\bf B}$ and $\Lambda \rightarrow \Lambda \times \left ({\bf B}^{-1} \right)^{\rm T}$ does not change TMI predictions. Thus, in order to directly compare TMI predictions with the ground truth, we need to reorient the TMI-inferred variables. To that end, we find a matrix ${\bf B}$ such that (1) $\bar Y_1$ and $\bar Y_2$ have the same dot product as the vectors $\bar E_{\rm int}$ and $\bar E_{\rm mag}$ and (2) $\bar Y_1$ is orthogonal to $\bar E_{\rm int}$.  In Fig.~\ref{fg:ising_comparison} panels (b) and (c) we show that the reoriented extensive variables $\bar Y_1$ and $\bar Y_2$ closely approximate the the true extensive variables $E_{\rm mag}$ and $E_{\rm int}$ respectively only from 50 sampled distributions. Notably, no symmetry or any other constraint was imposed on the inferred extensive variables.

\begin{figure}
       \begin{center}
               \includegraphics[scale=0.43]{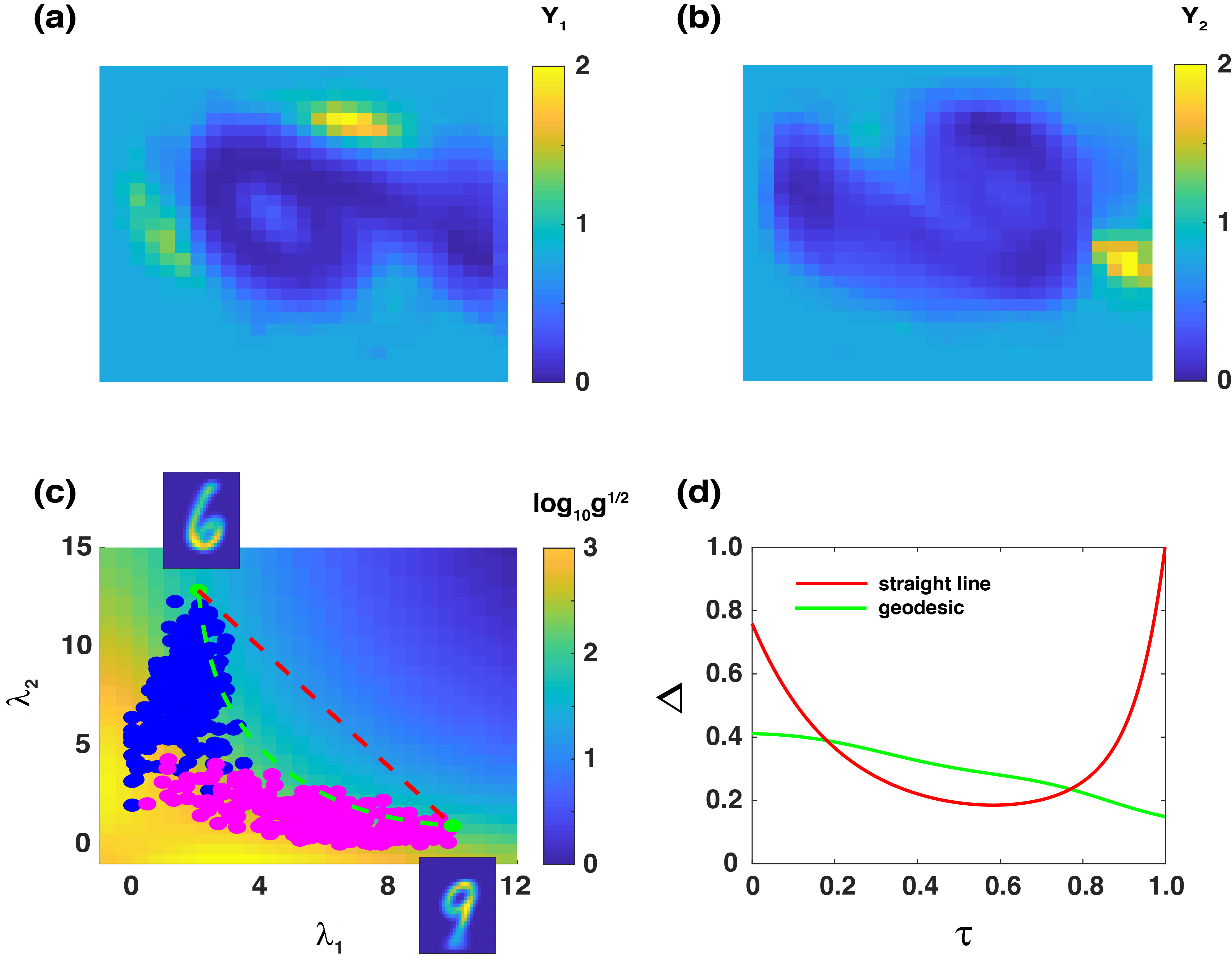}
               \caption{{\bf panel (a)} A heatmap of the inferred extensive variable $Y_1$ representing a generic `9'. {\bf panel (b)} A heatmap of the inferred extensive variable $Y_2$ representing a generic `6'. {\bf panel (c)} A scatter plot of the intensive bath parameters of the 500 data points. The data labeled `6' are colored blue while the data labeled `9' are colored magenta. The heatmap represents the volume element (square root of the deteminant of the metric tensor). The dashed red line is a straight line transformation between two data points shown at the top left and bottom right. The dashed green line is the geodesic computed using the Fisher-Rao metric. {\bf panel (d)} The symmetrized Kullback-Leibler divergence $\Delta$ between successive intermediate transformations along a discretization of the straight line trajectory (red) and the geodesic trajectory (green) of the transformation shown in panel (c).\label{fg:mnist}}
       \end{center}
\end{figure}

{\bf Analysis of handwritten digits:} We illustrate the application of TMI using the MNIST dataset~\citep{lecun2010mnist}.  We randomly selected 500 digits from the set of all `6's and `9's from MNIST. The digits were represented as a $28 \times 28$ array of positive numbers. Each data point was normalized and treated as a distribution represented by a 784 dimensional probability vector. Given that there were two types of digits,  we set out to infer $K=2$  sample-independent extensive variables. We simultaneously inferred  the corresponding intensive variables for individual data points. We imposed the positivity constraint on the intensive variables (see Appendix~\ref{ap_numerical}). In  panels (a) and (b) of Fig.~\ref{fg:mnist} we show the two inferred extensive variables $\bar Y_1$ and $\bar Y_2$. Notably, TMI correctly identifies two extensive variables (potential energy functions) that correspond to a generic digit `9' and a generic digit `6' respectively. These represent  the two {\it potential energy minima} in the data.

Moreover, as shown in panel (c), the two digits can also be classified by two different regions of the space of intensive variables; `9's are characterized by a high $\lambda_1$ and a low $\lambda_2$ while `6's are characterized by a low $\lambda_1$ and a high $\lambda_2$. Importantly, the Fisher-Rao metric on the space of intensive variables defines a notion of distance between the distributions as well as the ``number of points'' in any given volume element~\citep{caticha2015basics}.  The heatmap in panel (c) represents the logarithm of the volume element given by the square root of the determinant of the Fisher information matrix.  It is clear that the reduced dimensional space is highly inhomogeneus; the same small change in $\lambda_1$ and $\lambda_2$ may have very different effects on the resulting distributions depending on the region of the space. 

Finally, the Fisher-Rao metric  allows us to construct geodesics between pairs of data point. As shown in panel (c) of Fig.~\ref{fg:mnist}, the geodesic (dashed green line) between an `6' (green circle, top left) and a `9' (green circle, bttom right) is substantially different than the straight line (dashed pink line). The geodesic can be used to  perform a smooth transformation between the two distributions. For any transformation curve $\gamma(\tau)$, we can compute $\Delta_\gamma(\tau)$ as the symmetrized Kullback-Leibler divergence  between successive distributions along the curve. A uniform $\Delta$ implies a net transformation that is equally spread out over the entire trajectory. In contrast, a varying $\Delta$ implies a `rough' transformation. Interestingly, as shown in panel (d) of Fig.~\ref{fg:mnist},  the geodesic leads to a uniform$\Delta$ as opposed to the straight line transformation.

{\bf TMI outperforms NMF in data reconstruction and classification:}  We  compared the overall performance of TMI with a mathematically related technique, non-negative matrix factorization (NMF)~\citep{lee1999learning,hofmann1999probabilistic}. While TMI represents the thermodynamic potential of any state as a matrix product, NMF approximates the probabilities themselves as a matrix product. Briefly, in NMF, positive valued data is expressed as a product of two matrices:
\begin{eqnarray}
{\bf x}_{a}^{(\alpha)}  \approx q_{a}^{(\alpha)}  =  \sum_k {\bf l}_{k}^{(\alpha)}  {\bf y}_{ka}. \label{eq:nmf}
\end{eqnarray} 
The matrices ${\bf l}$ and ${\bf y}$ are determined by minimizing either the $L_2$ norm or the Kullback-Leibler divergence between the data $\{ {\bf x}_{a}^{(\alpha)}  \}$ and the approximation $\{ {q}_{a}^{(\alpha)}  \}$. NMF is a widely used technique to model positive valued data as it leads to interpretable positive valued decomposition (see~\citep{wang2012nonnegative} for a review). We note that NMF-based decomposition of the data is a linear superposition of positive valued `feature vectors' $\bar {\bf y}_k$s with positive valued `coefficients' $\bar {\bf l}^{(\alpha)} $s. In contrast, TMI expresses the data as a multiplicative decomposition (see Eq.~\ref{eq:gb}).

\begin{table}
\centering
\begin{tabular}{|c|cc|cc|cc|cc|}
\toprule
\hline
K &  \multicolumn{2}{c|}{MNIST} & \multicolumn{2}{c|}{Microbiome} & \multicolumn{2}{c|}{NIPS}& \multicolumn{2}{c|}{CBCL}\\
\hline
\midrule
{}   & TMI   & NMF    & TMI   & NMF & TMI   & NMF & TMI & NMF \\ \hline
1   &  0.89 & 0.92   & 0.29 & 0.30 & 0.24 & 0.24 & 0.033 & 0.036 \\
2   &  0.78  & 0.82   & 0.18  & 0.20 &0.22 & 0.23 & 0.022 & 0.028 \\
3   &  0.68  &  0.74   & 0.14  & 0.16&0.21 & 0.22  & 0.019 & 0.022\\
4   &  0.59 &  0.68   & 0.11  & 0.14&0.21 & 0.21 & 0.018 & 0.019 \\
5   &  0.53  &  0.64   & 0.09  & 0.12&0.20 & 0.21  & 0.016 & 0.017\\
10   &  0.34  &  0.50   & 0.05  & 0.07& 0.18 & 0.19  & 0.010 & 0.012\\
20   &  0.17  &  0.38  & 0.02  & 0.04& 0.15 & 0.18 & 0.005 & 0.007 \\
40   &  0.07  &  0.26   & 0.01  & 0.01 & 0.12 & 0.16  & 0.002 & 0.003 \\ \hline
\bottomrule
\end{tabular}
\caption{Comparison of KL divergences between the data $\{ {\bf x}_{a}^{(\alpha)}  \}$ and the approximate reconstruction $\{ q_{a}^{(\alpha)}  \}$ using TMI and NMF respectively. $K$ indicates the number of extensive variables used to model the data. The divergences are reported as an average per data point.}\label{tb:comparison}
\end{table}

To compare the ability of TMI and NMF to approximate the data, we chose four datasets of very different origins. The first was the MNIST dataset of handwritten digits~\citep{lecun2010mnist}. From the MNIST dataset, we randomly selected 500 samples comprising digits from $0$ to $9$. As above, each digit was represented by a $28\times 28$ array of pixel intensities which was normalized to 1. The second was the time series data collected on the gut microbiome of a human~\citep{david2014host}.  The microbiome data consisted of 318 samples collected approximately daily over a period of a year from the feces of one human individual. Each sample  was represented by the relative abundances of 70 most abundant bacterial operational taxonomic units (OTUs). The third dataset comprised a `bag of words' description~\citep{zhang2010understanding} of papers submitted to the Neural Information Processing Systems conference (downloaded from~\citep{asuncion2007uci}). Each paper was represented as a collection of words wherein each word was assigned a frequency in each submitted article. The fourth dataset comprised 472 grayscale images of human faces stored as an array of $19\times 19$ pixels (the CBCL database of faces~\citep{rowley98})  (see appendix~\ref{ap_data} for details of the datasets). 

We approximated each of the datasets  using TMI and NMF with several different values of $K$. For each $K$ we compared the Kullback-Leibler divergence between the data $\{ {\bf x}_{a}^{(\alpha)}  \}$ and the reconstruction $\{ q_{a}^{(\alpha)}  \}$. As seen in Table~\ref{tb:comparison}, TMI consistently performed better than NMF at reconstructing the data for every value of $K$. One possible reason behind this is that real data sets often have widely varying amplitudes. For example, the intensity of any given pixel in a set of images can vary substantially from image to image~\citep{ruderman1994statistics}. The exponential tuning of probabilities using the intensive variables  in TMI may be better suited to capture such variability compared to the linear superposition in NMF.  

Next, we tested how TMI performed in data classification using the MNIST dataset. To that end, used the 500 MNIST digits as above and inferred intensive variables  and extensive variables across a range of $K$ values. We used these intensive variables and the known identities of the digits to train a support vector machine (SVM) classifier. Next, we randomly selected 2000 digits from the dataset and predicted their identities.  Similarly, we fitted the same data with NMF and trained an SVM classifier with the same hyperparameters. The accuracy of the two identifications is shown in Table~\ref{tb:classify}. Similar to its ability to fit the data accurately, TMI also performs significantly better than NMF at classifying the data. 
\begin{table}
\centering
\begin{tabular}{|c|c|c|}
\toprule
\hline
K &  TMI & NMF \\
\hline
\midrule
5    &  $0.62 \pm 0.05$ & $0.49\pm 0.04$ \\
10   &   $0.75 \pm 0.04$&   $0.63 \pm 0.05$ \\
15   &  $ 0.77 \pm 0.03$ &  $0.67 \pm 0.05$ \\
20  &  $0.80 \pm 0.04$ & $0.69 \pm 0.05$\\ \hline
\bottomrule
\end{tabular}
\caption{Comparison of classification success rate (fraction) of randomly chosen handwritten digits from the MNIST dataset using TMI and NMF. The error bars are standard deviations estimated using 20 equal subsamples of the test set.}\label{tb:classify}
\end{table}

{\bf Discussion:} The manifold assumption~\citep{belkin2003laplacian}, commonly invoked in modern data analysis, posits that high dimensional data is generated by a few governing parameters and as a result can be represented by a lower dimensional manifold residing in the higher dimension. Several manifold inference methods such as diffusion maps~\citep{coifman2006diffusion},  Laplacian Eigenmaps~\citep{belkin2003laplacian}, Isomaps~\citep{balasubramanian2002isomap}, tSNE (t-stochastic neighorhood embedding)~\citep{maaten2008visualizing}, and UMAP (uniform manifold approximation and projection)~\citep{mcinnes2018umap} have been developed to approximately reconstruct these manifolds from the data. 

While the manifold-based methods achieve dimensionality reduction, unlike other approaches such as principal component analysis (PCA) or nonnegative matrix factorization (NMF)~\citep{lee1999learning,hofmann1999probabilistic}, they cannot obtain an approximate reconstruction of the original data using lower dimensional `features'. At the same time, these  methods do not obtain an analytical description of the manifold but only visualize it using a non-linear embedding of the data points in the lower dimensional space.  As a result, analytical manipulations such as computation of geodesics and volume elements are not possible. 

We presented TMI, an approach rooted in statistical physics to approximately embed positive valued high dimensional data points in lower dimensions. TMI possesses advantages of both manifold approximation methods as well as matrix-based dimensionality reduction methods. (1) similar to matrix-based methods such as PCA, SVD, and NMF, TMI can approximate data using lower dimensional features. Notably, similar to NMF, these features are positive valued (see Eq.~\ref{eq:gb}) and thus interpretable. Moreover, given the multiplicative nature of the decomposition, TMI appears to be better suited to model real data compared to NMF. (2) Similar to manifold approximation methods, TMI can infer an approximate lower dimensional manifold on which the data resides. Importantly, unlike previously developed methods (discussed above), TMI defines an analytically tractable and readily computable Riemannian manifold (with an associated distance metric) in the lower dimension. This in turn allows us to compute geodesics and volume elements in the reduced dimensional description. 

While TMI outperformed NMF in modeling and classifying data, in the current implementation, TMI was slower than NMF. Therefore, in the future, it will be important to optimize the numerical algorithms in TMI. Similarly, the calculation of the geodesic can be time consuming given that it requires solving boundary value non-linear differential equations. However, numerically efficient techniques have been developed~\citep{heymann2008pathways,heymann2008geometric,rotskoff2017geometric} which will be more useful in  situations when using $K > 2$ extensive variables. Another potential way to avoid solving the non-linear differential equations is to rely on the observation that the geodesics pass through the data rich regions of the $\bar \lambda$ space. Consequently, we can potentially approximate the geodesic as the shortest path on a graph connecting the data points themselves.

Finally, we comment on another potential approach to quantify differences between distributions using non-equilibrium statistical physics. The approach presented in this work relies on excess work in a nonadiabatic transformation which takes a system from a thermodynamic equilibrium with a bath $\bar \lambda^{(1)} $ to a thermodynamic equilibrium with the bath $\bar \lambda^{(2)} $. In contrast, we can also set up a system that is simulataneously in contact with the two baths $\bar \lambda_1$ and $\bar \lambda_2$. Such a system will reach a non-equilibrium steady state and will constantly dissipate heat from one bath to another. The steady state entropy production rate which is always positive and only zero when $\bar \lambda^{(1)}  = \bar \lambda^{(2)} $ can also be used as a quantifier of the differences between distributions. This rate is by definition positive and can be constructed to be symmetric. However, it remains to be seen whether it defines a distance metric. 

{\bf Acknowledgments:} We would like to thank Dr. Shreya Saxena for numerous discussions about the topic and Dr. Manas Rachh for useful comments on the manuscript. 
%

\newpage

\setcounter{figure}{0}
\makeatletter
\renewcommand{\thefigure}{A\@arabic\c@figure}
\makeatother

\setcounter{equation}{0}
\makeatletter
\renewcommand{\theequation}{A\@arabic\c@equation}
\makeatother

\setcounter{section}{0}
\makeatletter
\renewcommand{\thesection}{A\@arabic\c@section}
\makeatother

\section{Hessian with respect to $\lambda$s\label{ap_hess}}
We have the cost
\begin{eqnarray}
C &=& \sum_\alpha \log Z^{(\alpha)} + \sum_{\alpha, a, k}  {\bf x}_{a}^{(\alpha)}\lambda_{k}^{(\alpha)} Y_{ka} \nonumber \\ &+& \sum_k \beta_k \left ( \sum_a Y_{ka}^2 \right )
\end{eqnarray}

Let us  derive the Hessian with respect to $\lambda$s. We have the gradient:
\begin{eqnarray}
\frac{\partial C}{\partial \lambda_{k}^{(\alpha)}} = \sum_a {\bf x}_{a}^{(\alpha)}Y_{ka} -\sum_a q_{a}^{(\alpha)} Y_{ka}. \label{eq:grl}
\end{eqnarray}
Differentiating Eq.~\ref{eq:grl} with respect to $\lambda_{j}^{(\gamma)}$, we find that the Hessian is simply the covariance matrix:
\begin{eqnarray}
\frac{\partial^2 C}{\partial \lambda_{k}^{(\alpha)} \lambda_{j}^{(\gamma)}} = \delta_{\alpha \gamma}\left (\langle Y_{ka}Y_{ja}\rangle_{\alpha} - \langle Y_{ka} \rangle_{\alpha}\langle Y_{ja}\rangle_{\alpha}  \right ). \label{eq:hrl}
\end{eqnarray}
In Eq.~\ref{eq:hrl}, the angular brackets $\langle \cdot \rangle_{\alpha}$ denote an average with respect to the $\alpha^{\rm th}$ approximation:
\begin{eqnarray}
\langle \cdot \rangle_\alpha = \sum_a q_{a}^{(\alpha)} (\cdot)
\end{eqnarray}
and $\delta_{\alpha\gamma}$ is the Kronecker delta function.

\section{Cost function is convex in $\lambda$s and $Y$s\label{ap_convex}}

In this appendix, we show that the (1) cost function in Eq.~\ref{eq:cost_mod} is convex with respect to $\lambda_{k}^{(\alpha)}~\forall~k\in [1, K]$ when all the other $\lambda$s and all the $Y$s are fixed and (2) the cost function is convex in $Y_{ka}~\forall~a\in [1, d]$ when $\lambda$s and all other $Y$s are fixed. 

The double derivative of the cost function for a fixed $\alpha$ is given by Eq.~\ref{eq:hrl1}: 
\begin{eqnarray}
\frac{\partial^2 C}{\partial \lambda_{k}^{(\alpha)} \lambda_{j}^{(\alpha)}} = \left (\langle Y_{ka}Y_{ja}\rangle_{\alpha} - \langle Y_{ka} \rangle_{\alpha}\langle Y_{ja}\rangle_{\alpha}  \right ). \label{eq:hrl1}
\end{eqnarray}
The matrix in Eq.~\ref{eq:hrl1} is a covariance matrices. Given that covariance matrices are non-negative, the Hessian matrix in Eq.~\ref{eq:hrl1} is non-negative as well.

Next, we look at the Hessian with respect to the $Y_{ka}$s  for a fixed $k$ when $\lambda$s and other $Y$s are fixed. We have the derivative:
\begin{eqnarray}
\frac{\partial C}{\partial Y_{ka}} &=& -\sum_{\alpha} \lambda_{k}^{(\alpha)}q_{a}^{(\alpha)} + 2\beta_k Y_{ka} + \sum_{\alpha}  {\bf x}_{a}^{(\alpha)}\lambda_{k}^{(\alpha)} \label{eq:gry}
\end{eqnarray}

\begin{eqnarray}
\frac{\partial^2 C}{\partial Y_{ka}Y_{kb}} = 2\delta_{ab}\beta_k  -\sum_{\alpha} \lambda_{k}^{(\alpha)}\frac{\partial q_{a}^{(\alpha)}}{\partial Y_{kb}} .\label{eq:hry1}
\end{eqnarray}

We have the derivative,
\begin{eqnarray}
\frac{\partial q_{a}^{(\alpha)}}{\partial Y_{kb}} &=& \frac{\partial}{\partial Y_{kb}} \frac{f_{a}^{(\alpha)}}{Z^{(\alpha)}} \label{eq:inter}
\end{eqnarray}
where
\begin{eqnarray}
f_{a}^{(\alpha)} = \exp \left( -\sum_k \lambda_{k}^{(\alpha)} Y_{ka}\right ).
\end{eqnarray}
From Eq.~\ref{eq:inter}, we have
\begin{eqnarray}
\frac{\partial q_{a}^{(\alpha)}}{\partial Y_{kb}} = \frac{1}{Z^{(\alpha)}}\frac{\partial f_{a}^{(\alpha)}}{\partial Y_{kb}} - q_{a}^{(\alpha)}\frac{\partial \log Z^{(\alpha)}}{\partial Y_{kb}}
\end{eqnarray}
We have
\begin{eqnarray}
\frac{\partial \log Z^{(\alpha)}}{\partial Y_{kb}} = -\lambda_{k}^{(\alpha)}q_{b}^{(\alpha)}
\end{eqnarray}
and
\begin{eqnarray}
\frac{\partial f_{a}^{(\alpha)}}{\partial Y_{kb}} = -\delta_{ab}\lambda_{k}^{(\alpha)}f_{a}^{(\alpha)}
\end{eqnarray}
Putting everything together, we have
\begin{eqnarray}
\frac{\partial q_{a}^{(\alpha)}}{\partial Y_{kb}}  &=&  -\delta_{ab}q_{a}^{(\alpha)}\lambda_{k}^{(\alpha)}  + q_{a}^{(\alpha)}\lambda_{k}^{(\alpha)}q_{b}^{(\alpha)} \\
&=& q_{a}^{(\alpha)}\lambda_{k}^{(\alpha)} \left ( -\delta_{ab} + q_{b}^{(\alpha)} \right )
\end{eqnarray}
Thus, the elements of the Hessian are given by
\begin{eqnarray}
\frac{\partial^2 C}{\partial Y_{ka}Y_{kb}} &=& 2\delta_{ab}\beta_k  - \sum_\alpha \left ( \lambda_{k}^{(\alpha)}\right)^2q_{a}^{(\alpha)} \left (q_{b}^{(\alpha)} - \delta_{ab} \right ) \nonumber \\
&=& 2\delta_{ab}\left ( \beta_k + \frac{1}{2}\sum_\alpha \left (\lambda_{k}^{(\alpha)}\right)^2 q_{a}^{(\alpha)} \right ) \nonumber \\ &-& \sum_\alpha \left (\lambda_{k}^{(\alpha)}\right )^2q_{a}^{(\alpha)} q_{b}^{(\alpha)} \nonumber \label{eq:hry} \\
\end{eqnarray}
Given that $q_{a}^{(\alpha)} > q_{a}^{(\alpha)}q_{b}^{(\alpha)}$ and $\beta_k > 0$, the sum of off-diagonal entries in the Hessian matrix is smaller than the diagonal entry, according to Gershgorin's disc theorem, the Hessian matrix in Eq.~\ref{eq:hry} will be positive semidefinite.  

\section{Numerical algorithms for parameter inference\label{ap_numerical}}
We numerically the $\lambda$s and the $Y$s using a combination of gradient descent and Hessian descent. For a fixed value of $Y$s, the Hessian matrix $H$ with respect to the $\lambda$s is given by Eq.~\ref{eq:hrl} and the gradient with respect to $\lambda$s is given by
\begin{eqnarray}
gr(\bar \lambda^{(\alpha)})_k \equiv \frac{\partial C}{\partial \lambda_{k}^{(\alpha)}} = \sum_a x_{a}^{(\alpha)} Y_{ka} - \sum_{a} q_{a}^{(\alpha)}(\bar \lambda^{(\alpha)}) Y_{ka} \nonumber \\
\end{eqnarray}
The Hessian descent algorithm updates the Lagrange multipliers as
\begin{eqnarray}
\bar \lambda^{(\alpha)} \leftarrow \bar \lambda^{(\alpha)}  - \eta_\lambda H(\bar \lambda^{(\alpha)})^{-1}gr(\bar \lambda^{(\alpha)})
\end{eqnarray}
where $\eta_\lambda$ is a learning rate chosen between 0 and 0.05 and $H$ is the Hessian matrix. In a single $\lambda$-iteration, we update individual $\bar \lambda^{(\alpha)}$ for a fixed value of $Y$s. 

Similarly, the Hessian with respect to $Y$s is given by Eq.~\ref{eq:hry} and the gradient is given by Eq.~\ref{eq:gry}. The $Y$s are also updated for each $k$ individually using a Hessian descent scheme:
\begin{eqnarray}
\bar Y_{k} \leftarrow \bar Y_{k}  - \eta_Y H(\bar Y_k)^{-1}g(\bar Y_{k}).
\end{eqnarray}
The learning rate $\eta_Y$ is also chosen between 0 and 0.05. 

\subsection{Enforcing positivity in the inference}
Positivity constraints on $\lambda$s are enforced using a multiplicative gradient descent algorithm~\citep{lee2001algorithms}. The gradient with respect to $\lambda_{k}^{(\alpha)}$ is given by
\begin{eqnarray}
gr^{(\alpha)}_{k} = \sum_a {\bf x}_{a}^{(\alpha)} Y_{ka} - \sum_a q_{a}^{(\alpha)} Y_{ka}
\end{eqnarray}
Given that the TMI based predictions do not depend on a translational shift in the extensive variables $Y$s, when learning $\lambda$s, we frame-shift the extensive variables to ensure that all $Y$s are positive. Then, we identify
\begin{eqnarray}
g_{k}^{(\alpha)} = \left (gr_{k}^{(\alpha)} \right)^{+} - \left (gr_{k}^{(\alpha)} \right)^{-} 
\end{eqnarray}
where
\begin{eqnarray}
 \left (gr_{k}^{(\alpha)} \right)^{+}= \sum_a {\bf x}_{a}^{(\alpha)} Y_{ka} > 0
\end{eqnarray}
and 
\begin{eqnarray}
 \left (gr_{k}^{(\alpha)} \right)^{-} = \sum_a x_{a}^{(\alpha)} Y_{ka} > 0
\end{eqnarray}
are both positive. We start with positive valued $\lambda$s and update them using a multiplicative procedure~\citep{lee2001algorithms}:
\begin{eqnarray}
\lambda_{k}^{(\alpha)} \leftarrow \lambda_{k}^{(\alpha)} \cdot \left( \frac{ \left (gr_{k}^{(\alpha)} \right)^{-}}{ \left (gr_{k}^{(\alpha)} \right)^{+}} \right )^{\eta}
\end{eqnarray}
where $\eta > 0$ is a learning rate.

\section{TMI for unnormalized data\label{ap_unnormalized}}

In this section, we show how TMI can work for unnormalized data. The cost function for unnormalized distributions can be written as~\citep{lee2001algorithms}
\begin{eqnarray}
C = \sum_{a,\alpha} \left (x_{a}^{(\alpha)} \log \frac{x_{a}^{(\alpha)}}{q_{a}^{(\alpha)}} - x_{a}^{(\alpha)} + q_{a}^{(\alpha)} \right ) + \sum_k \beta_k \left ( \sum_a Y_{ka}^2 \right ) \nonumber \\
\end{eqnarray}
where
\begin{eqnarray}
q_{a}^{(\alpha)} = \exp \left( -\sum_{k=1}^{K} \lambda_{k}^{(\alpha)} Y_{ka} \right )
\end{eqnarray}
is the unnormalized distribution and $\{ x_{a}^{(\alpha)} \}$ is the unnormalized positive valued data. We rewrite $C$ after dropping terms that do not depend on $\lambda$s and $Y$s:
\begin{eqnarray}
C = \sum_{a,\alpha, k} x_{a}^{(\alpha)} \lambda_{k}^{(\alpha)}Y_{ka} + \sum_{a,\alpha} q_{a}^{(\alpha)} + \sum_k \beta_k \left ( \sum_a Y_{ka}^2 \right ) \nonumber \label{eq:cost_ap}  \\
\end{eqnarray}

We differentiate Eq.~\ref{eq:cost_ap} with respect to $\lambda_{k}^{(\alpha)}$ and set the derivative to zero:
\begin{eqnarray}
\sum_a x_{a}^{(\alpha)} Y_{ka} = \sum_a q_{a}^{(\alpha)} Y_{ka} \label{eq:grl_un}
\end{eqnarray}
Notably, Eq.~\ref{eq:grl_un} are identical to the normalized version (see Eq.~\ref{eq:dl}). Similarly, we differentiate with respect to $Y$s and set the gradient to zero:
\begin{eqnarray}
0 &=& \sum_{\alpha} x_{a}^{(\alpha)} \lambda{k}^{(\alpha)} + 2\beta_k Y_{ka} - \sum_\alpha q_{a}^{(\alpha)} \lambda_{k}^{(\alpha)} \\
\Rightarrow Y_{ka} &=& \frac{1}{2\beta_k} \sum_\alpha \left (\lambda_{k}^{(\alpha)} \left ( x_{a}^{(\alpha)} - q_{a}^{(\alpha)} \right ) \right ) \label{eq:gry_un}
\end{eqnarray}
Similar to Eq.~\ref{eq:grl_un}, Eq.~\ref{eq:gry_un} are identical to Eq.~\ref{eq:dy}. This indicates that the stationary points of the cost function do not depend on whether the data is normalized or not.

\section{Computing the friction tensor\label{ap_friction}}

Consider a transition rate matrix $\kappa$ whose stationary distribution is given by $q_a(\bar \lambda)$. The probability of being in state $b$ at time $t$ conditioned on being in state $a$ at time $t=0$ is given by $K_{ab}$ where the matrix $K$ is givenby
\begin{eqnarray}
K = \exp \left ( \kappa \tau \right ) = V \exp \left (\Lambda \tau \right ) V^{-1}.
\end{eqnarray}
where $V\Lambda V^{-1}$ is the diagonalization of $\kappa$. We can now express the friction tensor: 
\begin{eqnarray}
g_{ij}(\bar \lambda) &=& \int \limits_{0}^{\infty} \langle\delta Y_i (0) \delta Y_j(\tau) \rangle_{\bar \lambda} d\tau\\
&=& \int \limits_{0}^{\infty} \left (\sum_{a,b} q_a\delta Y_{ia}\delta Y_{jb}K_{ab} \right ) d\tau \\
&=& \int \limits_{0}^{\infty}  C_i \exp \left (\Lambda \tau\right ) D_jd\tau
\end{eqnarray}
where
\begin{eqnarray}
C_{i} = \left (q \circ \delta \bar Y_i  \right )^{\rm T} V~{\rm and}~D_j = V^{-1}\delta \bar Y_j
\end{eqnarray}
where $a\circ b$ is the Haddamard (elementwise) product. Thus, we have
\begin{eqnarray}
 g_{ij} &=&\int \limits_{0}^{\infty}  \sum_a C_{ia}D_{ja} \exp \left (\Lambda_a \tau\right )  d\tau\\
&=& -\sum_a \frac{C_{ia}D_{ja}}{\Lambda_a}
\end{eqnarray}
where the sum omits the zero eigenvalue. 
\section{Data for NMF/TMI comparison and implementation of NMF\label{ap_data}}

\subsection{Microbiome data}
The microbiome data was onbtained from David et al.~\citep{david2014host}. Briefly, the data consisted of bacterial operational taxonomic unit (OTU) abundances collected over a period of a year. There were 318 samples; each samples comprised relative abundances of $\sim 8\times 10^{3}$ OTUs. Based on our previous analysis~\citep{ji2019quantifying}, we discarded from the data OTUs whose average relative abundance was less than $0.1\%$ as these abundances are likely to represent technical noise in data collection. The data on remaining 70 high abundant OTUs was renormalized to relative fractions. 

\subsection{Bag of words data from NIPS conferences}

The bag of words description~\citep{zhang2010understanding} is a simple way to characterize text documents. Briefly, for a collection of documents, one first identifies all possible words. Next, the frequency of each word in each document is estimated. The document is then represented as a vector of frequencies, regardless of the order in which the words appear.  

We dowloaded the bag of words model of article submissions to the NIPS conference from the UCI machine learning repository~\citep{asuncion2007uci}. From the data,  we removed article submissions that were characterized by less than 1000 words and words that had less than 100 appearances across all articles. The resultant dataset had 1322 article each represented by a normalized probabilty vector with 2753 entries.

\subsection{Implementation of nonnegative matrix factorization}

We implemented a modified algorithm to learn the matrices {\bf l} and {\bf y} in Eq.~\ref{eq:nmf}. We followed the update algorithm that corresponds to minimization of Kullback-Leibler divergence between the data and the approximate representation~\citep{lee2001algorithms}. To ensure normalization of the approximate reconstruction, in each iteration of ${\bf l}$, for each $\alpha$, we multiplied the vectors $\bar{\bf l}^{(\alpha)}$  such that the predictions $q^{(\alpha)}$ sum to one.  

\section{Geodesic equations\label{ap_geodesic}}

Here, we explicitly write down the geodesic equations between two points in the space of intensive parameters when $K = 2$. The geodesics are calculated for the Fisher-Rao metric. This will facilitate numerical implementation of the Geodesic equations.

The metric tensor of the Fisher-Rao metric is given by
\begin{eqnarray}
G(i,j) =\begin{pmatrix}
\langle Y_i^2\rangle - \langle Y_i\rangle^2& \langle Y_iY_j\rangle - \langle Y_i\rangle \langle Y_j\rangle \\
\langle Y_iY_j\rangle - \langle Y_i\rangle \langle Y_j\rangle & \langle Y_j^2\rangle - \langle Y_j\rangle^2
\end{pmatrix} \nonumber \\
\end{eqnarray}
Let us denote by $F$ the inverse of $G$, $F = G^{-1}$. We note that $F$ is symmetric.

The first step towards writing the geodesic equations is estimating the Christoffel symbols. The Christoffel symbols of the first kind are given by~\citep{brody2008information}
\begin{eqnarray}
-2\Gamma_{ijk} &=&  -\frac{\partial^3 }{\partial \lambda_i \partial\lambda_j \partial\lambda_k} \log Z\\
 &=& \langle Y_iY_jY_k\rangle - \langle Y_iY_j \rangle\langle Y_k\rangle -  \langle Y_iY_k \rangle\langle Y_j\rangle \nonumber \\
&-&  \langle Y_kY_j \rangle\langle Y_i\rangle + 2\langle Y_i\rangle\langle Y_j\rangle\langle Y_k\rangle
\end{eqnarray}
Given the symmetry of the Fisher-Rao metric, there are only four unique Christoffel symbols of the first kind; $\Gamma_{111}, \Gamma_{112}, \Gamma_{122}$, and $\Gamma_{222}$. Moreover, the symbols do not change with permutation of the indices. They are
\begin{eqnarray}
-2\Gamma_{111} &=& \langle Y_1^3\rangle - 3\langle Y_1^2 \rangle\langle Y_1\rangle + 2\langle Y_1\rangle^3\\
-2\Gamma_{222}&=&\langle Y_2^3\rangle - 3\langle Y_2^2 \rangle\langle Y_2\rangle + 2\langle Y_2\rangle^3\\-2\Gamma_{112}&=&\langle Y_1^2Y_2\rangle - \langle Y_1^2 \rangle\langle Y_2\rangle -  2\langle Y_1Y_2 \rangle\langle Y_1\rangle \nonumber \\
&+& 2\langle Y_1\rangle^2\langle Y_2\rangle\\
-2\Gamma_{122} &=& \langle Y_2^2Y_1\rangle - \langle Y_2^2 \rangle\langle Y_1\rangle -  2\langle Y_1Y_2 \rangle\langle Y_2\rangle \nonumber \\
&+& 2\langle Y_1\rangle\langle Y_2\rangle^2
\end{eqnarray}

The Christoffel symbols of the second kind are given by
\begin{eqnarray}
\Gamma^{i}_{jk} =\sum_l F_{il} \Gamma_{ljk}.
\end{eqnarray}
Similar to the first kind, there are only 4 Christoffel symbols of the second kind. They are
\begin{eqnarray}
\Gamma^{1}_{11} &=& F_{11} \Gamma_{111} + F_{12} \Gamma_{211}\\
\Gamma^{2}_{22} &=& F_{21} \Gamma_{122} + F_{22} \Gamma_{222}\\
\Gamma^{1}_{12} &=& F_{11} \Gamma_{112} + F_{12} \Gamma_{212}\\
\Gamma^{1}_{22} &=& F_{11} \Gamma_{122} + F_{12} \Gamma_{222}
\end{eqnarray}

Finally, let us write down the geodesic differential equations~\citep{brody2008information}:
\begin{eqnarray}
\frac{d^2 \lambda_1}{dt^2} + \sum_{k,l}\Gamma^{1}_{kl}\frac{d\lambda_k}{dt}\frac{d\lambda_l}{dt} = 0\\
\frac{d^2 \lambda_2}{dt^2} + \sum_{k,l}\Gamma^{2}_{kl}\frac{d\lambda_k}{dt}\frac{d\lambda_l}{dt} = 0
\end{eqnarray}
Expanding further:
\begin{eqnarray}
0&=& \frac{d^2 \lambda_1}{dt^2} + \sum_{k,l}\Gamma^{1}_{kl}\frac{d\lambda_k}{dt}\frac{d\lambda_l}{dt} \\
&=&\frac{d^2 \lambda_1}{dt^2} + \Gamma^{1}_{11}\frac{d\lambda_1}{dt}^2 + \Gamma^{1}_{22}\frac{d\lambda_2}{dt}^2+ 2\Gamma^{1}_{12}\frac{d\lambda_1}{dt}\frac{d\lambda_2}{dt}  \nonumber \\
\end{eqnarray}
and
\begin{eqnarray}
0&=&\frac{d^2 \lambda_2}{dt^2} + \sum_{k,l}\Gamma^{2}_{kl}\frac{d\lambda_k}{dt}\frac{d\lambda_l}{dt} \\
&=&\frac{d^2 \lambda_2}{dt^2} + \Gamma^{2}_{11}\frac{d\lambda_1}{dt}^2 + \Gamma^{2}_{22} \frac{d\lambda_2}{dt}^2+ 2\Gamma^{2}_{12}\frac{d\lambda_1}{dt}\frac{d\lambda_2}{dt} \nonumber \\
\end{eqnarray}

\section{Figure for Ising model}
\begin{figure}[h]
\includegraphics[scale=0.5]{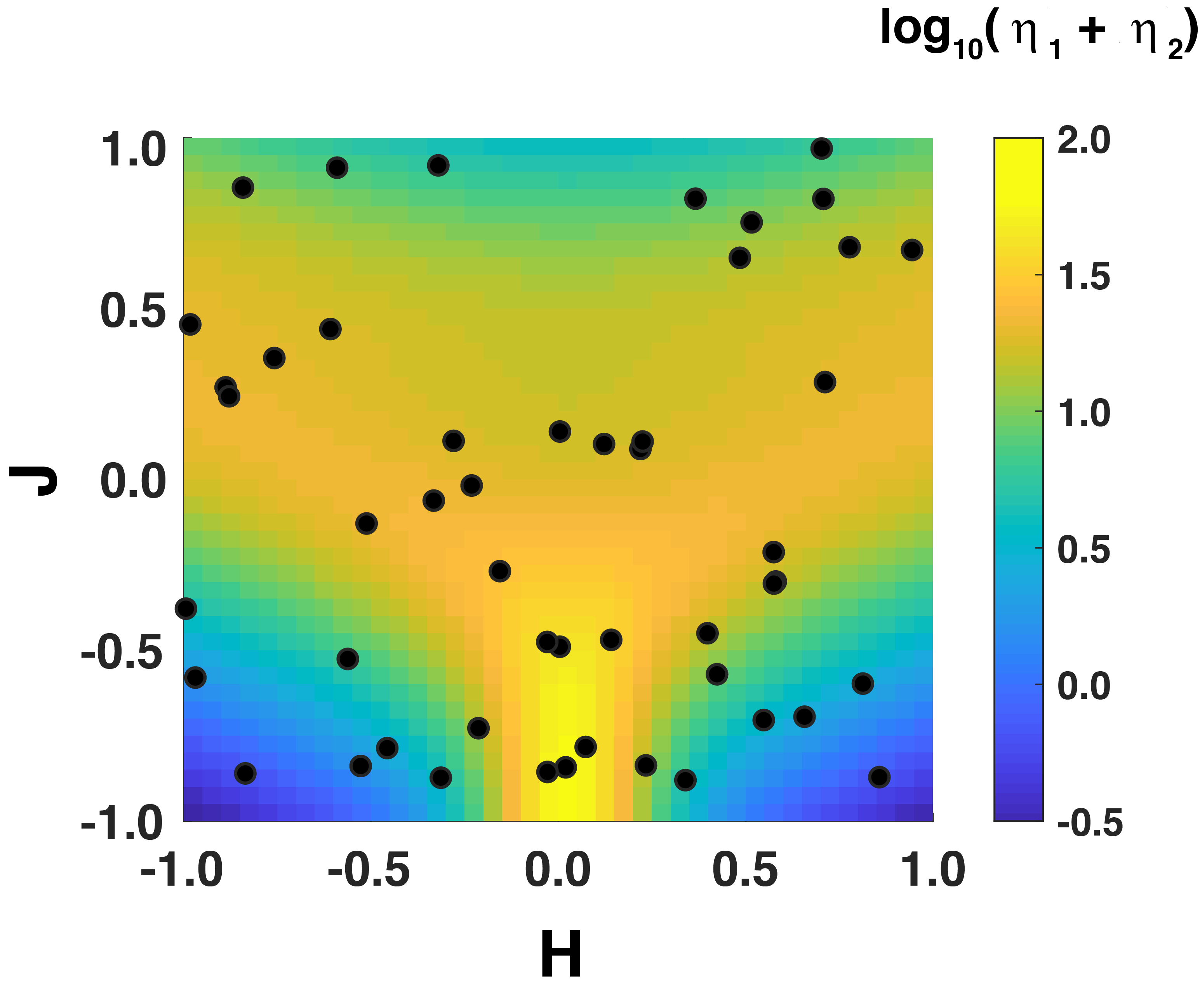}
\caption{Scatter plot of 50 randomly chosen $H$ and $J$ values between $[-1 1]$. The color represents the logarithm of the trace (sum of eigenvalues $\eta_1$ and $\eta_2$) of the Fisher information matrix of the Ising model~\citep{rotskoff2015optimal}. \label{fg:ap_ising}}
\end{figure}

\end{document}